\documentclass[%
 reprint,
superscriptaddress,
groupedaddress,
 amsmath,amssymb,
aps,
prl,
]{revtex4-1}
\usepackage{graphicx}  
\usepackage{dcolumn}   
\usepackage{bm}        
\usepackage{verbatim}  
\usepackage{sidecap}

\usepackage{physics}
\usepackage{float}
\usepackage{color}
\usepackage[dvipsnames]{xcolor}
\usepackage[colorlinks]{hyperref}

\newcounter{myequation}
\newcounter{myfigure}
\makeatletter
\@addtoreset{equation}{myequation}
\@addtoreset{figure}{myfigure}
\makeatother

\def \dag{^{\dagger}}
\def \ham{\mathcal{H}}
\def \S{\mathbf{S}}
\newcommand{\vect}[1]{\mathbf{\bm{#1}}}
\newcommand{\uvec}[1]{\hat{\vect{#1}}}

\begin{document}

\title{Mechanism for subgap optical conductivity in honeycomb Kitaev materials}
\author{Adrien Bolens}
\email{bolens@spin.phys.s.u-tokyo.ac.jp}
\author{Hosho Katsura}
\author{Masao Ogata}
\author{Seiji Miyashita}
\affiliation{
    Department of Physics, University of Tokyo, Hongo, Bunkyo-ku, Tokyo 113-0033, Japan
   }

\date{\today}
\begin{abstract}
Motivated by recent terahertz absorption measurements in $\alpha$-RuCl$_3$, we develop a theory for the electromagnetic absorption of materials described by the Kitaev model on the honeycomb lattice. We derive a mechanism for the polarization operator at {\it second order} in the nearest-neighbor hopping Hamiltonian. Using the exact results of the Kitaev honeycomb model, we then calculate the polarization dynamical correlation function corresponding to electric dipole transitions in addition to the spin dynamical correlation function corresponding to magnetic dipole transitions. 
\end{abstract}

\maketitle
\textit{Introduction}.
In Mott insulators, the electronic charge is localized at each site due to Coulomb repulsion, and the low-energy properties are described by the remaining spin and orbital degrees of freedom. Small charge fluctuations subsist due to virtual hopping of the electrons and generate the effective interaction. The same fluctuations can be responsible for a finite effective polarization operator \cite{bulaevskii2008electronic,khomskii2010spin,kamiya2012multiferroic,hwang2014signatures,batista2016frustration}. As a result, some magnetic systems can respond to an external ac electric field in a non-trivial way \cite{ng2007power,elsasser2012power,pilon2013spin,potter2013mechanisms}.
In the case of the simple single-band Hubbard model, such an effect has been calculated at third order in the virtual hopping, and is generally predicted only in frustrated lattices \cite{bulaevskii2008electronic,khomskii2010spin,batista2016frustration}.
 
In this Rapid Communication, motivated by recent experiments of terahertz spectroscopy of $\alpha$-RuCl$_3$ \cite{little2017antiferromagnetic,wang2017magnetic}, we consider the case of Kitaev materials: multi-orbital Mott insulators which are in close proximity to the Kitaev honeycomb model \cite{kitaev2006anyons}. We show that by introducing additional on-site degrees of freedom, the restriction to frustrated lattices can be lifted, and we derive an effective polarization operator on each bond of the lattice.

The Kitaev honeycomb model is exactly solvable and possesses a quantum spin liquid (QSL) ground state. Its potential realization in real materials through the Jackeli-Khaliullin mechanism \cite{jackeli2009mott} attracted much attention in recent years.
Exact analytical results for spin correlations have been derived for the Kitaev model \cite{baskaran2007exact} and used to predict the signatures of Majorana quasiparticles in inelastic neutron \cite{knolle2014dynamics,knolle2015dynamics,song2016low,knolle2016dynamics}, Raman \cite{knolle2014raman,knolle2016dynamics,nasu2016fermionic} and resonant x-ray scatterings \cite{halasz2016resonant}. As yet, potential Kitaev materials, such as Na$_2$IrO$_3$ \cite{chaloupka2010kitaev,jackeli2009mott,liu2011long,choi2012spin,ye2012direct} and $\alpha$-RuCl$_3$ \cite{pollini1996electronic,plumb2014alpha,kim2015kitaev,johnson2015monoclinic,sears2015magnetic,cao2016low}, all eventually reach a magnetically ordered state at sufficiently low temperatures \cite{liu2011long,choi2012spin,ye2012direct,johnson2015monoclinic,sears2015magnetic,cao2016low}, indicating significant deviations from the Kitaev model \cite{matsuura2014poor}. Nevertheless, in the case of $\alpha$-RuCl$_3$, experimental observations of a residual continuum of excitations have been interpreted as remnants of the Kitaev physics \cite{sandilands2015scattering,banerjee2016proximate,banerjee2017neutron,little2017antiferromagnetic}.

In the terahertz absorption measurement \cite{little2017antiferromagnetic}, it is argued that the absorption continuum of $\alpha$-RuCl$_3$ is too strong to be attributed to direct coupling to magnetic dipole (MD) moments, so that there must be a contribution from electric dipole (ED) transitions.
We study the response of low-energy excitations of Kitaev materials to an electromagnetic field by deriving a new microscopic mechanism.
We show that the interplay of Hund's coupling, spin orbit coupling (SOC) and a trigonal crystal field (CF) distortion results in a finite polarization operator up to \textit{second order} in the nearest neighbor hopping term, which is different from previous results obtained only at the third order. This is an important result as it sheds light on a new way to derive an electric polarization of pure electronic origin which is potentially relevant for various multi-orbital Mott insulators.
We then calculate the optical conductivity at $T=0$ in the ideal case of a pure Kitaev model by combining analytical and numerical methods. We thus show that the fractionalized low-energy excitations, although emerging from an effective spin Hamiltonian, respond to an external \textit{electric} field.

\textit{Model}. The Hamiltonian of Kitaev materials has been discussed extensively in the literature \cite{rau2014generic,rau2014trigonal,sizyuk2014importance,kim2016crystal,winter2016challenges,yadav2016kitaev,wang2017theoretical,winter2017models,hou2017unveiling}. The nearly octahedral ligand field strongly splits the $e_{g}$ orbitals from the $t_{2g}$ ones. For $d^5$ filling, one hole occupies the three $t_{2g}$ orbitals per site with an effective angular momentum $L=1$.
We thus study a tight-binding model for the holes,
\begin{equation}
\label{eq:fullham}
  \ham = \ham_{\rm hop} + \ham_{\rm SOC} + \ham_{\rm CF} + \ham_{\rm int}, 
\end{equation}
which is the sum of the kinetic hopping term, SOC, CF splitting among the $t_{2g}$ orbitals, and the Coulomb and Hund interactions, respectively.

In the present Rapid Communication, we consider Hamiltonians with the full $C_3$ symmetry (which may be appropriate for $\alpha$-RuCl$_3$ \cite{kim2016crystal, wang2017theoretical}). In addition, we only consider nearest neighbor hopping processes.
 
\begin{figure}
\centering
\includegraphics[width=0.3\textwidth]{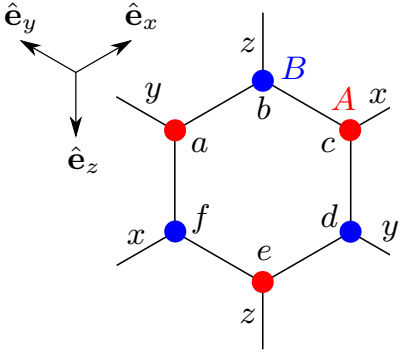}
	\caption{One hexagon of the honeycomb lattice. The different bond types ($x,y,z$) are indicated, along with their respective unit vectors $\uvec{e}_x$, $\uvec{e}_y$, and $\uvec{e}_z$. The $A$ and $B$ sublattices are colored in red and blue respectively.}
\label{fig:honeycomb}
\end{figure}

The Hamiltonians are concisely expressed by using the hole operators
\begin{equation}
  \mathbf{c}_i \dag=(c\dag_{i, yz,\uparrow},c\dag_{i, yz,\downarrow},c\dag_{i, xz,\uparrow},c\dag_{i, xz,\downarrow},c\dag_{i, xy,\uparrow},c\dag_{i, xy,\downarrow}).
\end{equation}
The kinetic term is $\ham_{\rm hop} = -\sum_{ \langle ij \rangle } \mathbf{c}_i\dag ( \mathbf{T}_{ij} \otimes \mathbb{I}_{2\times 2} ) \mathbf{c}_j$, where $\mathbb{I}_{2\times 2}$ is the $2\times 2$ identity matrix, and $\mathbf{T}_{ij}$'s are the hopping matrices among the $d_{yz}$, $d_{xz}$, and $d_{xy}$ orbitals,
\begin{equation}
\label{eq:hoppingmatrix}
  \mathbf{T}^x = \begin{pmatrix}
  t_3 & t_4 & t_4 \\
  t_4 & t_1 & t_2 \\
  t_4 & t_2 & t_1
\end{pmatrix}, 
\mathbf{T}^y = \begin{pmatrix}
  t_1 & t_4 & t_2 \\
  t_4 & t_3 & t_4 \\
  t_2 & t_4 & t_1
\end{pmatrix}, 
\mathbf{T}^z = \begin{pmatrix}
  t_1 & t_2 & t_4 \\
  t_2 & t_1 & t_4 \\
  t_4 & t_4 & t_3
\end{pmatrix},
\end{equation}
where $x,y$, and $z$ refer the type of the bond considered [see Fig.~\ref{fig:honeycomb}] and $t_{1-4}$ are the different hopping integrals. 
The SOC Hamiltonian is given by ${\cal H}_{\rm SOC} = \lambda/2 \sum_{i,a} {\bf c}^\dagger_i (L^a \otimes \sigma^a) {\bf c}_i $
with $\lambda >0$, where $(L^a)_{bc}= -i \epsilon_{abc}$ and $\sigma^a$ are the Pauli matrices.
The $C_3$ symmetric CF splitting of the $t_{2g}$ orbitals corresponds to a trigonal distortion along the axis perpendicular to the plane of the honeycomb lattices, $\ham_{\rm CF} = \Delta \sum_i \mathbf{c}_i\dag [ \qty(\mathbf{L}\cdot \uvec{n}_{\rm CF})^2 \otimes \mathbb{I}_{2\times 2}]\mathbf{c}_i$ with $\uvec{n}_{\rm CF} = [111]$.
The interaction Hamiltonian $\ham_{\rm int}$ is the Kanamori Hamiltonian \cite{kanamori1963electron,georges2013strong, perkins2014interplay} with intra-orbital Coulomb repulsion $U$, interorbital repulsion $U'=U - 2J_H$, and Hund's coupling $J_H$,
\begin{align}
\label{eq:kanamori}
  \ham_{\rm int} = &U \sum_{i,a} n_{i,a,\uparrow} n_{i,a,\downarrow} + (U' -J_H) \sum_{i, a<b, \sigma} n_{i,a,\sigma}n_{i,b,\sigma} \nonumber \\
  & + U'_{i,a\neq b} n_{i,a,\uparrow} n_{i,b,\downarrow} - J_H\sum_{i,a\neq b} c_{i,a,\uparrow} \dag c_{i,a,\downarrow} c_{i,b,\downarrow}\dag c_{i,b,\uparrow} \nonumber \\
  & + J_H \sum_{i,a\neq b} c_{i,a,\uparrow}\dag c_{i,a,\downarrow}\dag c_{i,b,\downarrow} c_{i,b,\uparrow}.
\end{align}

In the limit $\lambda \gg t^2/U$, it is well known  (see, e.g., Refs.~\cite{winter2016challenges, kim2016crystal}) that at second order perturbation theory in $\ham_{\rm hop}$, the effective Hamiltonian is the KH$\Gamma\Gamma'$ model which includes the Heisenberg (H) and anisotropic ($\Gamma$ and $\Gamma'$) interactions (not shown here) in addition to the Kitaev model (K) described by
\begin{equation}
  \ham_{K} = -4 J_K \sum_{\langle ij \rangle_{\gamma}}S_i^{\gamma} S_j^{\gamma},
\end{equation}
for the effective spins $1/2$. The trigonal distortion is usually small and we treat it as a perturbation ($\abs{\Delta} \ll \lambda$) unless stated otherwise.

\textit{Polarization}. The on-site Hamiltonian breaks the particle-hole symmetry as $\ham_{\rm SOC}$ and $\ham_{\rm CF}$ are both antisymmetric under the particle-hole transformation. Therefore, in contrast to the single-band Hubbard model \cite{bulaevskii2008electronic}, a finite polarization operator at second order in $\ham_{\rm hop}$ is not forbidden, even though the lattice is bipartite.	

In the atomic limit ($t_{1-4}=0$), the system has exactly one hole per site, i.e., $n_i = 1$ for all sites $i$, where $n_i = \sum_{\alpha=1}^6 n_{i\alpha}$ ($\alpha$ labels the six $t_{2g}$ states). The polarization operator measures the deviation from this configuration and is defined as
$
  \mathbf{P} = e\sum_i \mathbf{r}_i \delta n_i,
$
where $\delta n_i = n_i - 1$ and $\mathbf{r}_i$ is the position of the site $i$. Conservation of charge entails $\sum_i \delta n_i = 0$. In the following, we set $e=1$.

We find the existence of a finite effective polarization operator at the second order in perturbation theory in $\ham_{\rm hop}$ if $\Delta$ and $J_H$ are finite. The effective polarization can be written as
\begin{align}
\label{eq:polbond}
  \mathbf{P}_{\rm eff} = \sum_{\langle ij \rangle_{\gamma}} \left( P_{ij} - P_{ji} \right) \uvec{e}_{\gamma} \equiv \sum_{\langle ij \rangle_{\gamma}} P_{ \langle ij \rangle} \uvec{e}_{\gamma},
\end{align}
where $\uvec{e}_{\gamma}$ is the unit vector along the $\gamma$ bond connecting the sites $i \in A$ sublattice and $j \in B$ sublattice [see Fig.~\ref{fig:honeycomb}], and $P_{ij}$ is given by perturbation theory. 
We can furthermore use the symmetry group of the bond $\langle ij \rangle$ to narrow down the possible terms in $P_{ \langle ij \rangle}$ \cite{miyahara2016theory}. Due to the hexagonal CF with the additional trigonal distortion, the symmetry group of a $\gamma$ bond of the honeycomb lattice is $\{e, i, C_2(\gamma), m_{\perp}(\gamma) \}$ whose elements are the identity element, the inversion transformation, the $C_2$ rotation around $\uvec{e}_{\gamma}$, and the reflection relative to the plane perpendicular to $\uvec{e}_{\gamma}$, respectively. Considering how $S_i^{a}S_j^{b}\uvec{e}_{\gamma}$ ($a,b \in \{x,y,z\}$) transforms under the different group elements, Eq.~(\ref{eq:polbond}) reduces to
\begin{equation}
\label{eq:poleff}
  \mathbf{P}_{\rm eff} = \sqrt{2}\mathbb{A} \sum_{\langle ij \rangle_{\gamma}}\left[ \uvec{e}_{\gamma} \cdot \left( \S_i \times \S_j \right) \right] \uvec{e}_{\gamma}.
\end{equation}
In the basis fixed by the octahedral CF (in which the Kitaev Hamiltonian is written), $\uvec{e}_{x}= (0,1,-1)/\sqrt{2}$, $\uvec{e}_{y}= (-1,0,1)/\sqrt{2}$, and $\uvec{e}_{z}= (1,-1,0)/\sqrt{2}$. Equation~(\ref{eq:poleff}) is valid for any general real symmetric hopping matrices which preserve the $C_3$ symmetry.
The unitless constant $\mathbb{A}$ is calculated at second order in $\ham_{\rm hop}$ by using the eigenstates of the $15\times 15$ two-hole on-site Hamiltonian. In order to obtain an analytical result, we only keep terms linear in $\Delta$. For $t_1 = t_3 = t_4 = 0$, we find
\begin{equation}
  \label{eq:Aconst}
  \mathbb{A} =  t_2^2 \Delta \left[ \frac{128(21 \lambda +8 U)}{81 \lambda  (3 \lambda +2 U)^4}  J_H  +O(J_H^2) \right],
\end{equation}
which scales as $t_2^2 \Delta  J_H/(U^3 \lambda)$ for $U \gg J_H, \lambda \gg \Delta$. The full expression (exact in $J_H$), and the expression including all the hopping integrals $t_{1-4}\ne 0$, is included in the Supplemental Material \cite{SM} together with details about the perturbation theory and numerical calculations of $\mathbb{A}$ exact in $\Delta$. 

Only a few hopping processes are possible at second order in $\ham_{\rm hop}$ (see the Supplemental Material \cite{SM}). Even when $\Delta =0$ or $J_H=0$, different allowed processes contribute to $P_{\langle ij \rangle}$, but they interfere destructively, and their contributions overall vanish. The interference is not completely destructive only when both $\Delta \neq 0$ and $J_H\neq 0$.

\textit{Optical conductivity}. The spin dynamics of the pure Kitaev model have been investigated thoroughly. However, for Kitaev materials, additional integrability breaking terms are indispensable. In particular, they explain the magnetic ordering at low temperature.
In this case, even the calculation of the spin structure factor becomes a challenge. Very recent works indicate that the spin dynamics evolve smoothly from the results of the pure Kitaev model using numerical \cite{winter2017breakdown,gohlke2017dynamics,gotfryd2017phase} and parton mean-field methods \cite{knolle2018dynamics} close to the QSL regime. In the following, we limit ourselves to the pure Kitaev model to calculate the polarization dynamical response of the fractionalized excitations, expecting that our results are meaningful physically in the putative proximate Kitaev spin liquids. We show that in this limit, the polarization dynamics is remarkably similar to the spin dynamics. The pure Kitaev limit is obtained from the electronic Hamiltonian by keeping only the $90^{\circ}$ metal-ligand-metal hoppings ($t_1=t_3=t_4=0$). Then, for small trigonal CF, the spin Hamiltonian becomes $\ham_{K} + O(\Delta)$. $\mathbf{P}_{\rm eff}$ is itself linear in $\Delta$, therefore we do not need the $O(\Delta)$ correction in the Hamiltonian to calculate the response at first order in $\Delta$.

The optical conductivity along the arbitrary in-plane direction $\uvec{e}_{\alpha}$ at $T=0$ for $\omega >0$ is 
\begin{align}
\label{eq:OC}
  \sigma^{\alpha}(\omega) = \frac{\omega}{V} {\rm Re} \Big \{\int_0^{\infty} dt e^{i\omega t}\langle P^{\alpha}(t)P^{\alpha}(0)\rangle \Big \},
\end{align}
where $\mathbf{P}(t) = e^{i\ham t} \mathbf{P} e^{-i\ham t}$, $P^{\alpha} = \mathbf{P} \cdot \uvec{e}_{\alpha}$ and $V$ is the volume of the system.
In the effective Kitaev model, we can substitute
$ \langle P^{\alpha}(t) P^{\alpha}(0) \rangle_{\rm Hubbard} \rightarrow \langle P_{\rm eff}^{\alpha}(t) P_{\rm eff}^{\alpha}(0) \rangle_{\rm Kitaev}
$ with $\mathbf{P}_{\rm eff}(t) = e^{i\ham_K t} \mathbf{P}_{\rm eff} e^{-i\ham_K t}$,
where the expectation value is taken with respect to the ground state of the Kitaev Hamiltonian. 

For the calculation of $\langle P^{\alpha}_{\rm eff}(t) P^{\alpha}_{\rm eff}(0) \rangle$, we need to evaluate spin correlations of the form $\langle S^a_i(t) S^b_j(t) S^c_k S^d_l\rangle$ for pairs of bonds $\langle ij \rangle$ and $\langle kl \rangle$, and for $a,b,c,d \in \{ x,y,z \}$. This is reminiscent of Raman scattering in the Kitaev Hamiltonian \cite{knolle2014raman}. However, unlike Raman scattering for which only terms with $a=b$ and $c=d$ are relevant, only terms with $a\neq b$ and $c \neq d$ terms appear in $\langle P^{\alpha}_{\rm eff}(t) P^{\alpha}_{\rm eff}(0) \rangle$, due to the anti-symmetric nature of $\mathbf{P}_{\rm eff}$. Moreover, we must have either $a=\gamma_{ij}$ or $b=\gamma_{ij}$, and similarly either $c=\gamma_{kl}$ or $d=\gamma_{kl}$.

On each hexagon, a specific product of Pauli matrices $W_p = 2^6 S_a^y S_b^z S_c^x S_d^y S_e^z　S_f^x$ (refer to Fig.~\ref{fig:honeycomb} for site labels) commutes with the Hamiltonian and has eigenvalues $\pm 1$, so that there is a conserved $\mathbb{Z}_2$ flux in each hexagon. Kitaev introduced an enlarged Hilbert space of Majorana fermions \cite{kitaev2006anyons}, in which the spin operators read
$  2 \hat{S}_i^a = i \hat c_i \hat b_i^a$.
We use a hat symbol to indicate that the operators act on the enlarged Hilbert space. The Majorana fermions $\hat c_i$ and  $\hat b_i^a$ are the matter and gauge fermions, respectively. The Kitaev Hamiltonian in terms of Majorana fermions is given by
\begin{equation}
\label{eq:hamMF}
  \hat{\ham}_{K}= i J_K \sum_{ \langle ij \rangle_{a} } \hat{u}_{\langle ij\rangle_{a}} \hat c_i \hat c_j, \qquad \hat{u}_{\langle ij\rangle_{a}} = i \hat b_i^a \hat b_j^a,
\end{equation}
where $\hat{u}_{\langle ij\rangle_{a}} = \pm 1 $ are constants of motions which fix the $\mathbb{Z}_2$ flux $W_p$ in each hexagon. For a fixed flux pattern, the remaining matter Hamiltonian is quadratic and thus solvable. We further introduce the bond fermions, which are complex fermions defined by ($i \in A$, $j \in B$)
\begin{align}
  \hat \chi_{\langle ij \rangle_{a}} = \frac{1}{2} (\hat b_i^a + i \hat b_j^a ), \qquad a = x,y,z.
\end{align}
In terms of the bond fermions, the spin operators become
\begin{align}
  \hat{S}^a_i  &= i \hat c_i(\hat \chi_{\langle ij \rangle_{a}} + \hat \chi\dag_{\langle ij \rangle_{a}})/2 \quad (i\in A) \nonumber \\
  &= \hat c_i(\hat \chi_{\langle ij \rangle_{a}} - \hat \chi\dag_{\langle ij \rangle_{a}})/2 \quad (i\in B).
\end{align}
In addition to adding a Majorana matter fermion at site $i$, $\hat{S}^a_i$ changes the bond fermion number of the bond $\langle ij \rangle_{a}$, which corresponds to the change $\hat{u}_{\langle ij \rangle_{a}} \rightarrow - \hat{u}_{\langle ij \rangle_{a}}$. Therefore, $\hat{S}^a_i$ adds one $\pi$ flux to the two plaquettes sharing the bond $\langle ij \rangle_{a}$, which is also true for $\hat{S}^a_i(t)$, at all times $t$, since all $\hat{u}_{\langle ij \rangle_{a}}$'s are constants of motion.

We now have a good criterion to identify which $\langle S^a_i(t) S^b_j(t) S^c_k S^d_l\rangle$ terms contribute to the optical conductivity.
The expectation value of a product of any operator can be finite only if it does not change the flux in any hexagons. This is a direct consequence of the orthogonality of the subspaces with different flux patterns. Therefore, only combinations of four spin operators which overall leave the flux in each hexagon unchanged are relevant.

For a fixed bond $\langle ij \rangle$, each pair $S^a_i S^b_j$ in $\mathbf{P}_{\rm eff}$ changes the fluxes in two adjacent hexagons. There are four different possible pairs of hexagons, located around the bond $\langle ij \rangle$.
\begin{figure}[htb]
\centering
\includegraphics[width=0.50\textwidth]{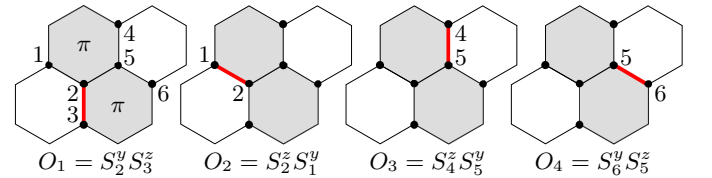}
	\caption{Four different operators $O_i$ appearing in $\mathbf{P}_{\rm eff}$ which create the same given pair of adjacent $\pi$ fluxes. The shaded hexagons represent the $\pi$ fluxes (plaquettes with $W_p=-1$).}
\label{fig:fixhexagons}
\end{figure}
Inversely, a fixed pair of adjacent hexagons is affected by four different operators $S^a_i S^b_j$. Let us consider the situation depicted in Fig.~\ref{fig:fixhexagons}. The four aforementioned operators are labeled $O_{1-4}$.
The different symmetries of the Kitaev model (mirror symmetries, $C_3$ symmetry and inversion symmetry) leave us with only four independent correlation functions
$\Gamma_i(t) = 2^4 \langle O_1(t) O_i(0) \rangle $, for $i = 1-4$,
from which we can calculate the full response,
\begin{equation}
  \sigma(\omega) = \frac{e^2 \mathbb{A}^2}{\hbar a_{\perp}} \frac{\omega}{8\sqrt{3}} ( 2 \tilde{\Gamma}_1 - 2 \tilde{\Gamma}_3 -  \tilde{\Gamma}_2 + \tilde{\Gamma}_4 ),
\end{equation}
where $\tilde{\Gamma}_i(\omega) = {\rm Re}\{\int_0^{\infty} dt e^{i\omega t}\Gamma_i(t)\}$. The calculated optical conductivity is independent of the direction of $\uvec{e}_{\alpha}$ and therefore isotropic on the plane.

Two different matter Hamiltonians are needed to calculate $\Gamma_{1-4}$ (see the Supplemental Material \cite{SM}): the flux-free Hamiltonian $\hat{\ham}_0$ (all $\hat{u}_{\langle ij \rangle}\textrm{'s} = 1$) and the two-flux Hamiltonian $\hat{\ham}'$, whose $\pi$ fluxes correspond to those depicted in Fig.~\ref{fig:fixhexagons} (all $\hat{u}_{\langle ij \rangle}\textrm{'s} = 1$ except for $\hat{u}_{\langle 25 \rangle} = -1$).
The ground-state of the full Kitaev Hamiltonian is in the flux-free sector \cite{lieb1994flux} so that it is given by the ground state $\ket{M_0}$ of $\hat{\ham}_0$.
Interestingly, the calculation of the simpler spin-spin dynamical correlation functions requires the same Hamiltonian $\hat{\ham}'$ \cite{baskaran2007exact, knolle2014dynamics}, implying that the magnetic dipole and electric dipole transitions take place between the same flux sectors.
Using the Lehmann spectral representation, we generally define the operator,
\begin{equation}
\label{eq:Piop}
  \Pi_{ab}(\omega) = \pi \sum_{\lambda} \langle M_0 | c_a | \lambda \rangle \langle \lambda | c_b | M_0 \rangle \delta(\omega - \Delta_{\lambda}),
\end{equation}
where $\ket{\lambda}$'s are the eigenvectors of $\hat{\ham}'$ with energy $E_{\lambda}$, and $\Delta_{\lambda} = E_{\lambda} - E_0$. We find
\begin{align}
\label{eq:gammas}
  \tilde{\Gamma}_1 = \Pi_{33}, \,\,\, \tilde{\Gamma}_2 = -\Pi_{31}, \,\,\, \tilde{\Gamma}_3 = i\Pi_{34}, \,\,\, \tilde{\Gamma}_4 = -i\Pi_{36},
\end{align}
where the matter fermions are labeled according to Fig.~\ref{fig:fixhexagons}. The imaginary part of the spin susceptibility can be written as $\chi''(\omega) \propto (\Pi_{22} + \Pi_{55} - i \Pi_{25} + i \Pi_{52})$ \cite{knolle2014dynamics}.
Using complex matter fermions and the appropriate Bogoliubov transformations, the Hamiltonians become
\begin{align}
\label{eq:diagham}
  \hat{\ham}_0 = \sum_{n>0} \omega_n a_n\dag a_n + E_0, \, \, \,
  \hat{\ham}' = \sum_{n>0} \omega'_n b_n\dag b_n + E'_0,
\end{align}
where $E_0 = - \frac{1}{2}\sum_{n>0} \omega_n$ and $E_0' = - \frac{1}{2}\sum_{n>0} \omega'_n$. Even though all the $\ket{\lambda}$ states are defined in the same flux sector, we find that the states that contributes to $\sigma(\omega)$ and $\chi''(\omega)$ are mutually exclusive. This can be explained using symmetries of the Hamiltonians $\hat{\ham}_0$ and $\hat{\ham}'$ \cite{SM, blaizot1986quantum}.

\begin{figure}
\centering
	\includegraphics[width=0.5\textwidth]{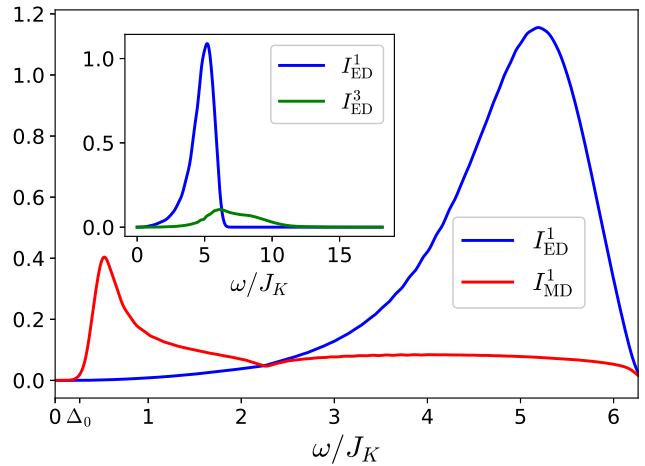}
	\caption{Single-particle responses $I^1_{\rm ED} = \sigma(\omega)$, in units of $e^2 \mathbb{A}^2/(\hbar a_{\perp})$, and $I^1_{\rm MD}$ where we set $R=(\mathbb{A}/ g)^2 \times (a/ \lambdabar)^2$ to $0.5$. The inset: single- and three-particle responses $I^1_{\rm ED}$ and $I^3_{\rm ED}$.}
\label{fig:OC}
\end{figure}

The electric dipole and magnetic dipole response functions were numerically calculated in systems of sizes up to $82 \times 82$ unit cells. As mentioned in Refs.~\cite{knolle2014raman, knolle2014dynamics, knolle2016dynamics}, the leading contribution comes from the single-particle states $\ket{\lambda} = b_{\lambda}\dag\ket{M'}$, where $\ket{M'}$ is the ground state of $\ham'$ which satisfies $b_{\lambda} \ket{M'} = 0$. We verified this property by calculating the single- and three-particle responses of a $20 \times 20$ system (see the Supplemental Material \cite{SM}), shown in the inset of Fig.~\ref{fig:OC}.

The electric and magnetic dipole absorption rates scale as $\sigma(\omega)$ and $\omega (g\mu_B/c)^2 \chi''(\omega)$ respectively, where $\mu_B$ is the Bohr magneton and $g$ is the effective Land\'e $g$-factor so that we set $I_{\rm ED}=\sigma(\omega)$ and $I_{\rm MD} = \omega (g\mu_B/c)^2 \chi''(\omega)$. The intensities are related by the ratio $R = (\mathbb{A}/ g)^2 \times (a/ \lambdabar)^2$, where $a$ is the spacing between the transition metal atoms on the honeycombs, and $\lambdabar$ is the reduced Compton wavelength. In the literature, a wide range of values for the different physical parameters has been reported, resulting in different values for $\mathbb{A}$. For $\alpha$-RuCl$_3$, the ratio $R$ roughly ranges between $0.01$ and $10$. Figure~\ref{fig:OC} shows $I_{\rm ED}$ in units of $e^2 \mathbb{A}^2/(\hbar a_{\perp})$ and the corresponding $I_{\rm MD}$ where the ratio $R$ is arbitrarily set to $0.5$. Around $\omega = \Delta_0$, $I_{\rm MD}$ seems to be dominant. However, $I_{\rm MD}$ is of course independent of $\mathbb{A}$, and its calculated value (with $g=2$) can only account for about $7 \%$ of the measured signal just above the sharp gap in Ref.~\cite{little2017antiferromagnetic}. With a ratio of $R = 10$, the order of magnitude of $I_{\rm ED}$ is comparable to the measured quantity, but the sharp gap at $\Delta_0$ disappears as $I_{\rm MD}$ becomes negligible.

\textit{Discussion}. We showed that the complex interplay of the Hund's coupling, SOC and a trigonal CF distortion results in a nontrivial polarization operator originating from nearest-neighbor hopping processes, shedding some light on an unexpected charge fluctuation mechanism in Kitaev materials. By calculating the effective polarization operator and its dynamical correlation function, we determined the electric dipole absorption spectrum originating from the pure Kitaev model. This shows that, like other spin liquids with a continuum of low-energy excitations \cite{ng2007power,elsasser2012power,pilon2013spin,potter2013mechanisms}, the fractionalized magnetic excitations respond to an external ac electric field.
As measured in the terahertz absorption measurements of $\alpha$-RuCl$_3$ \cite{little2017antiferromagnetic}, the electric dipole spectral weight is expected to dominate over the magnetic dipole one.
Our results for the optical conductivity in Fig.~\ref{fig:OC} are valid for the pure Kitaev model. However, the derived polarization operator (\ref{eq:poleff}) is valid even in the effective KH$\Gamma\Gamma'$ model (only the coefficient $\mathbb{A}$ is affected). Therefore, we expect the optical response to be modified smoothly when introducing integrability breaking terms in the proximity of the QSL regime, as for the spin structure factor \cite{winter2017breakdown,gohlke2017dynamics,gotfryd2017phase,knolle2018dynamics}. Nonetheless, substantial changes in the spin Hamiltonian, such as a large $\Gamma$ term (expected in real materials), most probably significantly alter the calculated response \cite{song2016low}.

Additionally, other corrections can potentially affect the optical conductivity in real materials, such as longer range hopping or breaking of the $C_3$ symmetry, which should explain the dependence on the direction of the probing ac field.

\textit{Acknowledgments}. A. B. thanks K. Penc and C. Hotta for helpful comments, and acknowledges FMSP for the encouragement of the present Rapid Communication. H. K. was supported, in part, by JSPS KAKENHI Grant No. JP15K17719 and No. JP16H00985. The present Rapid Communication was supported by the Elements Strategy Initiative Center for Magnetic Materials (ESICMM) under the outsourcing project of MEXT.


%

\widetext
\begin{center}
\pagebreak
\textbf{\large Supplemental Material of "Mechanism for subgap optical conductivity in honeycomb Kitaev materials"}
\end{center}
\stepcounter{myequation}
\stepcounter{myfigure}
\setcounter{table}{0}
\setcounter{page}{1}
\makeatletter
\renewcommand{\theequation}{S\arabic{equation}}
\renewcommand{\thefigure}{S\arabic{figure}}
\renewcommand{\bibnumfmt}[1]{[S#1]}
\section{Perturbation theory}
First we derive the expression for the effective polarization Eq.~(\ref{eq:poleff}) using second order perturbation in $\ham_{\rm hop}$ and treating all other Hamiltonian exactly. We also show explicitly the different hopping processes involved.

In the unperturbed Hilbert space of an $N$-site system, with $\ham_{\rm hop} = 0$, the $2^N$ degenerate eigenstates are the magnetic states $\ket{\phi_n}$, with exactly one hole per site. The perturbation lift the degeneracy and the new $2^N$ low-energy eigenstates $\ket{\psi_n}$ are adiabatically connected to the magnetic states, such that there are connected by a unitary transformation: $\ket{\psi_i} = e^{-S} \ket{\phi_i}$, where $S$ is antihermitian.

For any observable $\mathcal{O}$ defined in the full Hilbert space, an effective low-energy operator $\mathcal{O}_{\rm eff}$ can be defined by projecting in the subspace spanned by $\{ \ket{\psi_n} \}$: $\mathcal{O}_{\rm eff} = \mathbb{P}_{\phi} e^{S} \mathcal{O} e^{-S} \mathbb{P}_{\phi}$, where $\mathbb{P}_{\phi}$ is the projection operator onto the unperturbed magnetic Hilbert space. An equivalent definition in terms of individual matrix element is
\begin{equation}
  \mel{\phi_m}{\mathcal{O}_{\rm eff}}{\phi_n} \equiv  \mel{\psi_m}{\mathcal{O}}{\psi_n},
\end{equation}
which we use to calculate the effective polarization operator.

Without the trigonal distortion $\Delta$, the magnetic states are split into effective $J= 1/2$ and $3/2$ states \cite{jackeli2009mott, chaloupka2010kitaev, sizyuk2014importance, winter2016challenges} that we denote $\ket{J,M_J}$. They are given by ($X=yz$, $Y=xz$, $Z=xy$)
\begin{align}
\label{eq:states}
  \ket{\frac{1}{2},\frac{1}{2}} &= \frac{1}{\sqrt{3}} \qty(-\ket{d_X, \downarrow} -i \ket{d_Y, \downarrow} - \ket{d_Z, \uparrow}) \nonumber \\
  \ket{\frac{1}{2},-\frac{1}{2}} &= \frac{1}{\sqrt{3}} \qty( -\ket{d_X, \uparrow} +i \ket{d_Y, \uparrow} + \ket{d_Z, \downarrow}) \nonumber \\
  \ket{\frac{3}{2},\frac{3}{2}} &= \frac{1}{\sqrt{2}} \qty(-\ket{d_X, \uparrow} - i \ket{d_Y, \uparrow} ) \nonumber \\
  \ket{\frac{3}{2},\frac{1}{2}} &= \frac{1}{\sqrt{6}} \qty(-\ket{d_X, \downarrow} -i \ket{d_Y, \downarrow} + 2\ket{d_Z, \uparrow}) \nonumber \\
  \ket{\frac{3}{2},-\frac{1}{2}} &= \frac{1}{\sqrt{6}} \qty(\ket{d_X, \uparrow} -i \ket{d_Y, \uparrow} + 2\ket{d_Z, \downarrow}) \nonumber \\
  \ket{\frac{3}{2},-\frac{3}{2}} &= \frac{1}{\sqrt{2}} \qty(\ket{d_X, \downarrow} -i \ket{d_Y, \downarrow} ).
\end{align}
Note that the relative sign between the states is not always consistent in the literature. It is not relevant when only interested in the Hamiltonian, but it is for the polarization operator. We consistently choose the states such that $J^- \ket{1/2,1/2} = \ket{1/2,-1/2}$, and similarly for the $J=3/2$ states.
The different hopping processes appearing in the perturbation theory at second order in $\ham_{\rm hop}$ can then be written schematically using those states. However, we want to track the same processes when $\Delta \neq 0$. As the $e_g$ orbitals have been cast away, the quantization axis in Eq.~(\ref{eq:states}) is arbitrary. It is usually chosen to be the $z$ axis of the octahedral environment but it does not need to. If the CF distortion direction coincides with the quantization axis of $M_J$, the structure of Eq.~(\ref{eq:states}) is left unchanged, even though $J$ and $M_J$ are no longer good quantum numbers. The eigenstates can still be labeled $\phi_{M}^J$ such that $\phi_{M}^J \rightarrow \ket{J,M}$ when $\Delta \rightarrow 0$. They are given by (see \cite{jackeli2009mott, perkins2014interplay})
\begin{align}
\label{eq:statesCF}
  \phi^{1/2}_{1/2} &= \frac{1}{\sqrt{2}} \cos\theta \qty(-\ket{d_X, \downarrow} -i \ket{d_Y, \downarrow}) - \sin\theta \ket{d_Z, \uparrow} \nonumber \\
  \phi^{1/2}_{-1/2} &= \frac{1}{\sqrt{2}} \cos\theta \qty( -\ket{d_X, \uparrow} +i \ket{d_Y, \uparrow}) + \sin\theta \ket{d_Z, \downarrow} \nonumber \\
  \phi^{3/2}_{3/2} &= \frac{1}{\sqrt{2}} \qty(-\ket{d_X, \uparrow} - i \ket{d_Y, \uparrow} ) \nonumber \\
  \phi^{3/2}_{1/2} &= \frac{1}{\sqrt{2}} \sin\theta \qty(-\ket{d_X, \downarrow} -i \ket{d_Y, \downarrow}) + \cos\theta \ket{d_Z, \uparrow} \nonumber \\
  \phi^{3/2}_{-1/2} &= \frac{1}{\sqrt{2}} \sin\theta \qty(\ket{d_X, \uparrow} -i \ket{d_Y, \uparrow}) + \cos\theta \ket{d_Z, \downarrow} \nonumber \\
  \phi^{3/2}_{-3/2} &= \frac{1}{\sqrt{2}} \qty(\ket{d_X, \downarrow} -i \ket{d_Y, \downarrow} ),
\end{align}
where $\tan(2\theta) = 2\sqrt{2}\frac{\lambda}{\lambda-2\Delta}$. When $\Delta =0$, $\sin\theta = 1/\sqrt{3}$. The block-diagonal structure of the Kanamori Hamiltonian~(4) is also left invariant \cite{perkins2014interplay}, so that the different hopping processes are the same with and without $\Delta$.
For Kitaev materials, the CF distortion direction is $\uvec{n}_{\rm CF} = [111]$. We thus rotate the orbital and spin angular momentum so that $[111] \rightarrow [001]$, corresponding to the SU(2) unitary transformation $U$ and the SO(3) rotation $R$. The pay-off is that the hopping matrices now become $\mathbf{T}^{\gamma} \rightarrow R \mathbf{T}^{\gamma} R^{-1}$. 
Note that the final expression with respect to the effective spin operators has to be rotated back to the original frame to be consistent with Eq.~(\ref{eq:poleff}).

\begin{SCfigure}
\centering
	\includegraphics[width=0.49\textwidth]{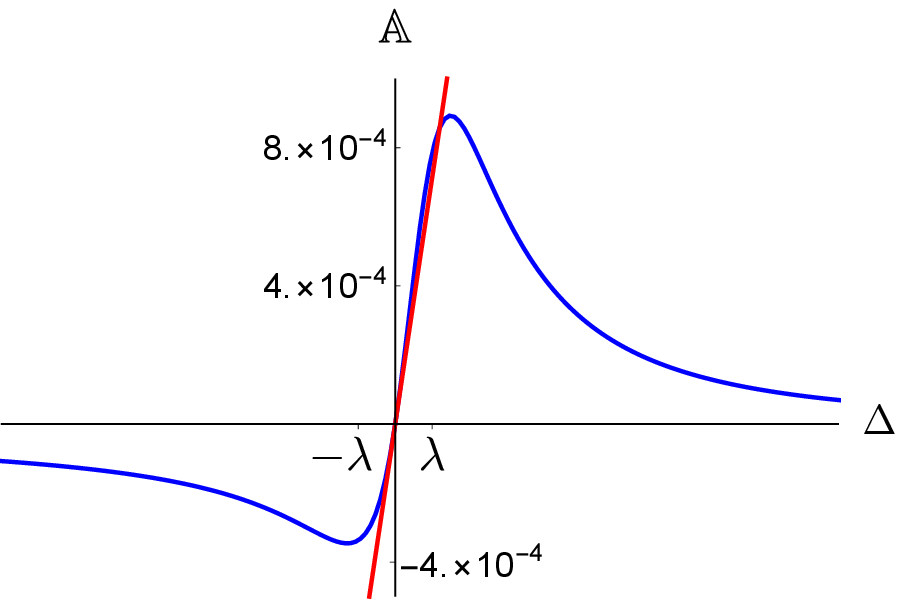}
	\caption{Coefficient $\mathbb{A}$ as a function of $\Delta$. Red line: analytical result linear in $\Delta$. Blue line: calculation exact in $\Delta$. $U=3000$, $J_H = 600$, $\lambda = 150$, $t = 100$.}
\label{fig:Acoef}
\end{SCfigure}

The exact eigenstates of the full Hamiltonian can always be decomposed in a magnetic state and a polar state (with one or more doubly occupied site): $\ket{\psi_m} = \alpha \ket{\phi_m} + \beta \ket{p_m}$, where $\alpha \simeq 1$. Only the polar states are relevant for the polarization operator, $\mel{\psi_m}{\mathbf{P}}{\psi_n} \propto \mel{p_m}{\mathbf{P}}{p_n}$. To calculate $\mathbf{P}_{\rm eff}$ at second order in $\ham_{\rm hop}$, all we need are the polar states at first order in $\ham_{\rm hop}$,
\begin{equation}
  \ket{p_m}_1 = \sum_{p \in {\rm polar}} \frac{\mel{p}{\ham_{\rm hop}}{\phi_m}}{E_0 - E_p} \ket{p},
\end{equation}
from which we deduce Eq.~(\ref{eq:polbond}) of the main text and
\begin{equation}
\label{eq:pol2nd}
  P_{ij} = \mathbb{P} \ham^{ij}_{\rm hop} \mathbb{Q} \frac{1}{(E_0 - \ham_0)^2} \mathbb{Q} \ham^{ji}_{\rm hop} \mathbb{P}.
\end{equation}
where $\ham_0$ is the unperturbed Hamiltonian with ground state energy $E_0$, and $\mathbb{P}$ and $\mathbb{Q}$ denote the projection operators onto the low-energy subspace made of effective $1/2$-spins, and the polar states (with doubly occupied sites), respectively.
The analytical calculation is too heavy, so that we only calculated the $\mathbb{A}$ coefficient numerically, shown as a function of $\Delta$ in Fig.~\ref{fig:Acoef}.

In order to obtain an analytical expression, we treat both $\ham_{\rm hop}$ and $\ham_{\rm CF}$ as perturbations. The ground state is thus constituted of the pure $J = 1/2$ states~(\ref{eq:states}) (the rotation $\uvec{n}_{\rm CF} \rightarrow [001]$ is unnecessary).

Up to third order, only terms scaling as $O(t^2 \Delta)$ are relevant and we need to calculate polar states at first order in both $\ham_{\rm hop}$ and $\ham_{\rm CF}$. We obtain
\begin{align}
\label{eq:pol3rd}
  P_{ij} = \left[-\frac{2}{3\lambda} \mathbb{P}_{\frac{1}{2}} \ham^{ij}_{\rm hop} \frac{\mathbb{Q}}{(E_0 - \ham_0)^2} \ham^{ji}_{\rm hop}  \mathbb{P}_{\frac{3}{2}}  \ham_{\rm CF} \mathbb{P}_{\frac{1}{2}} +  \mathbb{P}_{\frac{1}{2}} \ham^{ij}_{\rm hop} \frac{\mathbb{Q}}{(E_0 - \ham_0)^2} \ham_{\rm CF} \frac{\mathbb{Q}}{E_0 - \ham_0} \ham^{ji}_{\rm hop}   \mathbb{P}_{\frac{1}{2}} \right] + \rm{H.c.},
\end{align}
where $\mathbb{P}_{\frac{1}{2}}$, $\mathbb{P}_{\frac{3}{2}}$ and $\mathbb{Q}$ are the projection operators on the $J =1/2$ states, the $J = 3/2$ states, and the polar states, respectively. 
The full expression of Eq.~(\ref{eq:Aconst}) of the main text is $\mathbb{A} = \frac{128}{81} \Delta t^2 J_H\frac{P}{Q}$ with
\begin{align}
  P = & -16 (J_H-U) (2 J_H-U) (3 J_H-U)^3 + 2 \lambda  (U-3 J_H)^2 \left(379 J_H^2-404 J_H U+105 U^2\right) \nonumber \\
  & -3 \lambda ^2 (3 J_H-U) \left(1847 J_H^2-1650 J_H U+363 U^2\right)  +3 \lambda ^3 \left(6004 J_H^2-4743 J_H U+927 U^2\right) \nonumber \\
  & +27 \lambda ^4 (129U-334 J_H) +1701 \lambda ^5 \nonumber \\
   Q = & \lambda 
   (-6 J_H+3 \lambda +2 U)^2 \left(6 J_H^2-J_H (17 \lambda +8 U)+(3 \lambda +U) (3
   \lambda +2 U)\right)^3.
\end{align}
For the more general hopping matrices which preserve the $C_3$ symmetry,
\begin{equation}
\label{eq:genhopmatrix}
  \mathbf{T}^x = \begin{pmatrix}
  t_3 & t_4 & t_4 \\
  t_4 & t_1 & t_2 \\
  t_4 & t_2 & t_1
\end{pmatrix}, 
\mathbf{T}^y = \begin{pmatrix}
  t_1 & t_4 & t_2 \\
  t_4 & t_3 & t_4 \\
  t_2 & t_4 & t_1
\end{pmatrix}, 
\mathbf{T}^z = \begin{pmatrix}
  t_1 & t_2 & t_4 \\
  t_2 & t_1 & t_4 \\
  t_4 & t_4 & t_3
\end{pmatrix},
\end{equation}
we find
\begin{equation}
  \label{eq:Aconstgeneral}
  \mathbb{A} =  \Delta (t_1+t_2-t_3-t_4) \qty[ \frac{128}{81} \frac{8 U^3
   (t_1+t_2+t_3)+3 \lambda  U^2 (11 t_1+7 t_2+9 t_3) +24 \lambda ^2 U (2 t_1+t_3) +9 \lambda ^3 (2t_1+t_3)}{ \lambda  U^2 (3 \lambda +2
   U)^4} J_H  +O(J_H^2) ].
\end{equation}
The full expression can also be written as a fraction of polynomials but we do not write it explicitly. We see that even with more general hopping matrices, the effective polarization vanishes when either $J_H=0$ or $\Delta=0$. (This is still true with hopping matrices breaking the $C_3$ symmetry.) Numerical calculations show that when $t_1+t_2 -t_3 -t_4 = 0$ the polarization vanishes, even when treating $\Delta$ exactly.

\begin{figure}
\centering
\includegraphics[width=0.85\textwidth]{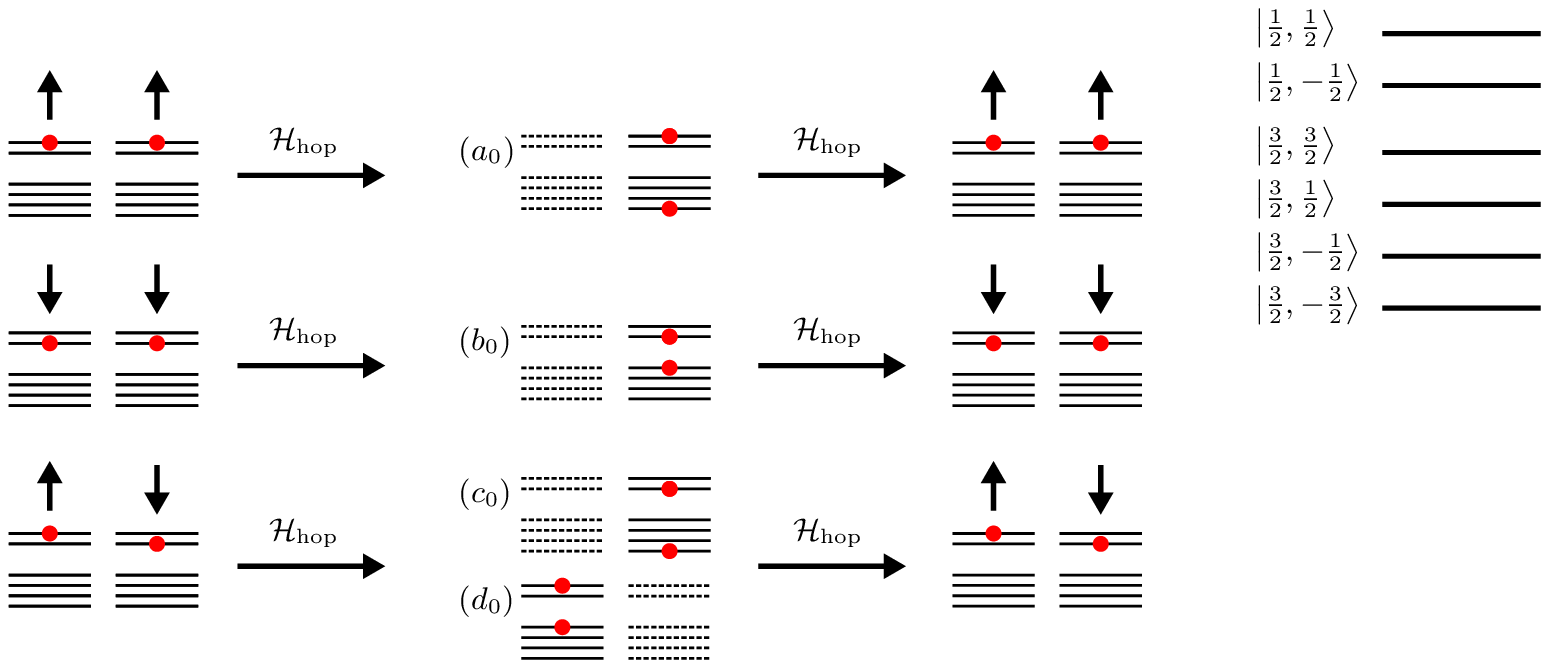}
	\caption{Different allowed hopping processes for a $z$ bond without the trigonal CF distortion, leading to the Kitaev Hamiltonian. Here, $M_J$ is quantized along the usual $z$ axis $[001]$ of the octahedral environment (in which the Kitaev Hamiltonian is written) as $\Delta =0$.}
\label{fig:pathsKitaev}
\end{figure}

Let us now consider explicitly the hopping processes. Without the CF distortion $\Delta$, the processes leading to the Kitaev Hamiltonian are handily visualized by choosing the usual quantization axis of $M_J$ (along $[001]$) and considering the hopping matrices with only $t_2\neq 0$. The only four possible processes for a $z$-bond are shown in Fig.~\ref{fig:pathsKitaev}, up to the exchange of the two sites. The mechanism is Ising-like as no spin flips are possible \cite{matsuura2014poor}. The Hund's coupling $J_H$ is responsible for the Ising ferromagnetism ($J_K > 0$). When $J_H=0$, the effective Hamiltonian is proportional to the identity ($J_K =0$).

Due to spatial inversion symmetry, only transitions between the singlet state and a triplet state are relevant to the effective polarization. With $\Delta =0$, the processes $(c_0)$ and $(d_0)$ actually individually allow such transitions, between the singlet and the $M_J=0$ triplet. However, they interfere destructively, resulting in a vanishing $\mathbf{P}_{\rm eff}$.

We now consider the more delicate case $\Delta \neq 0$. Therefore, we switch to the states~(\ref{eq:statesCF}), for which $M_J$ is quantized along $[111]$. We will only consider transitions between the triplet state $\ket{\uparrow, \uparrow}$ and the singlet state (if the polarization is considered) or the $M_J=0$ triplet state (if the Hamiltonian is considered).

\begin{figure}
\centering
\includegraphics[width=0.85\textwidth]{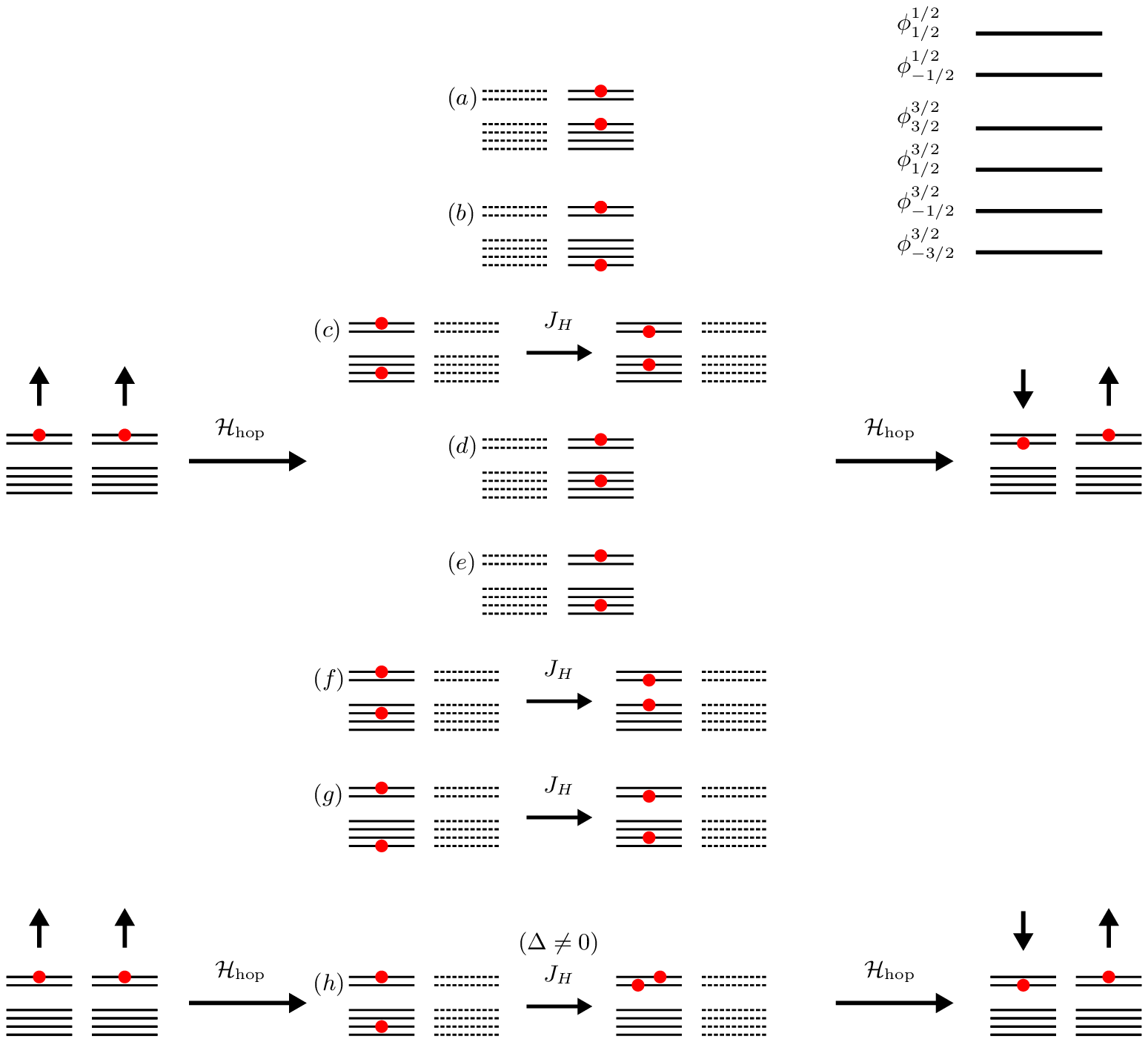}
	\caption{Different allowed hopping processes. The process $(h)$ is only possible when $\Delta \neq 0$. Here $M_J$ is quantized along $\uvec{n}_{\rm CF} = [111]$ and not the $z$ axis $[001]$ of the octahedral environment.}
\label{fig:paths}
\end{figure}

The different processes are represented graphically in Fig.~\ref{fig:paths}. Note that as the hopping matrices are rotated, no additional processes appear when considering the more general hopping matrices of Eq.~(\ref{eq:genhopmatrix}). As before, when $\Delta=0$, the processes $(a)$-$(g)$ of Fig.~\ref{fig:paths} interfere destructively. However, when both $\Delta \neq 0$ and $J_H \neq 0$, the processes $(a)$-$(g)$ do not completely cancel and $\mathbf{P}_{\rm eff}$ is finite. Te process $(h)$ is only possible when $\Delta \neq 0$ and $J_H \neq 0$ and also contribute to $\mathbf{P}_{\rm eff}$ (though its relative amplitude is small compared to the other processes).

\section{4-spin dynamical correlation function of the Kitaev Hamiltonian}

Here we derive the expressions for the different correlation functions $\Gamma_i$ given in Eqs.~(\ref{eq:Piop}) and (\ref{eq:gammas}). In the Majorana representation,
\begin{align}
  \Gamma_1(t) &= \bra{M_0}\bra{F_0} \sigma_2^y(t) \sigma_3^z(t) \sigma_2^y(0) \sigma_3^z(0) \ket{F_0}\ket{M_0} \nonumber \\
  \Gamma_2(t) &= \bra{M_0}\bra{F_0} \sigma_2^y(t) \sigma_3^z(t) \sigma_2^z(0) \sigma_1^y(0) \ket{F_0}\ket{M_0}  \nonumber \\
  \Gamma_3(t) &= \bra{M_0}\bra{F_0} \sigma_2^y(t) \sigma_3^z(t) \sigma_4^z(0) \sigma_5^y(0) P \ket{F_0}\ket{M_0}  \nonumber \\
  \Gamma_4(t) &= \bra{M_0}\bra{F_0} \sigma_2^y(t) \sigma_3^z(t) \sigma_6^y(0) \sigma_5^z(0) P \ket{F_0}\ket{M_0},
\end{align}
where $\ket{F_0}\ket{M_0}$ is the ground state of the Kitaev Hamiltonian $\ham_{K}$ decomposed into the gauge and matter sectors, such that $\hat{u}_{\langle ij \rangle} \ket{F_0} = \ket{F_0}$. $P$ is the projector onto the physical Hilbert space, defined by $D_j \ket{\Psi}_{\rm phys} = \ket{\Psi}_{\rm phys}$ where $D_j = c_j b_j^x b_j^y b_j^z$. It can be shown that $P$ is only needed when some of the bond fermion number operators $\chi_{\langle ij \rangle_{a}}\dag \chi_{\langle ij \rangle_{a}} = (\hat{u}_{\langle ij \rangle_a} +1)/2$ are not conserved (even though the $\mathbb{Z}_2$ fluxes are). This is the case for $\Gamma_3$ and $\Gamma_4$. Note that $P$ commutes with the spin operators.

The general strategy is to calculate separately the expectation value in the gauge sector and the matter sector. In terms of Majorana fermions,
\begin{align}
  \Gamma_1 = \bra{M_0}\bra{F_0}  e^{i\ham_0 t} ( i c_2 \chi_{21} \dag )  ( - c_3 \chi_{23} \dag ) e^{-i\ham t} ( i c_2 \chi_{21}  ) ( c_3 \chi_{23} )\ket{F_0}\ket{M_0},
\end{align}
where $\ham_0$ refers to the matter fermion Hamiltonian with all $u_{\langle ij\rangle_{a}} = 1$.
We now use the important relation
\begin{equation}
  \chi_{\langle ij \rangle_{\gamma}} \dag e^{-i\ham t} =  e^{-i(\ham + V_{ij}) t}  \chi_{\langle ij \rangle_{\gamma}} \dag,
\end{equation}
where $V_{ij} = -2i J_K c_i c_j$. This implies that $\ham_0 + V_{ij}$ is the Hamiltonian with all bond operators $\hat{u} = 1$ except on the bond $\langle ij \rangle_{\gamma}$ where $u_{\langle ij \rangle_{\gamma}} = -1$. Therefore,
\begin{align}
  \Gamma_1 = &\bra{M_0} e^{i\ham_0 t} c_2 c_3 e^{-i(\ham_0 + V_{21} + V_{23}) t} c_2 c_3 \ket{M_0} \bra{F_0}  \chi_{21} \dag  \chi_{23} \dag  \chi_{21}  \chi_{23} \ket{F_0} \nonumber \\
  =& -\bra{M_0} e^{i\ham_0 t} c_2 c_3 e^{-i(\ham_0 + V_{21} + V_{23}) t} c_2 c_3 \ket{M_0}
\end{align}

For the Kitaev Hamiltonian in a general flux sectors characterized by the set $\{ u_{\langle ij \rangle_{\gamma}} \}$, noted $\ham(\{ u\})$, we have the relation
\begin{equation}
  c_i \ham(\{ u\}) c_i = \ham(\{ \tilde{u}\}), \qquad  {\rm where} \quad \tilde{u}_{\langle mn \rangle} = \left\{
                \begin{array}{ll}
                  -u_{\langle mn \rangle} \quad {\rm if} \, i = m \, {\rm or}\,  n\\
                  u_{\langle mn \rangle} \quad \, \, \, \, \, {\rm else }
                \end{array}
              \right. .
\end{equation}
Together with the relation $c_i e^O c_i = e^{c_i O c_i}$, we have
\begin{align}
  \Gamma_1 = \bra{M_0} e^{i\ham_0 t} c_3 e^{-i\ham ' t} c_3 \ket{M_0}.
\end{align}
with $\ham ' = \ham_0 + V_{25}$.
We could further simplify $c_3 e^{-i\ham ' t} c_3 = e^{-i\ham '' t}$, but we do not for the following reason. 
$\ham_0$, $\ham '$ and $\ham ''$ are all Hamiltonians in the matter Hilbert space with a fixed bond fermion parity. 
However, the matter parities of their respecting ground states do not necessarily match. For example, in the general Kitaev Hamiltonian with different parameters for the three bonds ($J_K^X$, $J_K^Y$ and $J_K^Z$), the parity of the ground state in a fixed gauge sector depends on the values of the three parameters (see Ref.~\cite{knolle2014dynamics}).
Numerically, we find that the ground state of $\ham_0$ ($\ket{M_0}$) has the same parity as that of $\ham '$ ($\ket{M'}$) and that the ground state of $\ham ''$ ($\ket{M''}$) has the opposite parity so that $\bra{M_0}\ket{M''}=0$. For this reason, we work with $\ham '$ so that we can find a relation between $\ket{M_0}$ and $\ket{M'}$ explicitly, which will then be used in a Bogoliubov transformation.

For $\Gamma_2$ we similarly find 
\begin{align}
  \Gamma_2 = -\bra{M_0} e^{i\ham_0 t} c_3 e^{-i\ham ' t} c_1 \ket{M_0}.
\end{align}
For $\Gamma_3$ and $\Gamma_4$ we need to add the projection operator $P$.
Here, it is enough to replace $P$ with $D_2 D_5$ \cite{knolle2014raman, knolle2016dynamics} , which reads
\begin{align}
  D_2 D_5 &= -i u_{\langle 25 \rangle_{x}} c_2 c_5 (\chi_{21} \dag  + \chi_{21}) (\chi_{23} \dag  + \chi_{23})(\chi_{65} \dag  - \chi_{65})(\chi_{45} \dag  - \chi_{45}).
\end{align}
Therefore, 
\begin{align}
  \Gamma_3 =& i\bra{M_0} e^{i\ham_0 t} c_2 c_3 e^{-i(\ham_0 + V_{21} + V_{23}) t} c_4 c_5 c_2 c_5 \ket{M_0}  \bra{F_0}  \chi_{21}\dag \chi_{23}\dag \chi_{45}\dag \chi_{65}\dag \chi_{21} \chi_{23} \chi_{65} \chi_{45} \ket{F_0} \nonumber \\
  =& -i\bra{M_0} e^{i\ham_0 t} c_2 c_3 e^{-i(\ham_0 + V_{21} + V_{23}) t} c_2 c_4 \ket{M_0} \nonumber \\
  =& i\bra{M_0} e^{i\ham_0 t} c_3 e^{-i \ham ' t} c_4 \ket{M_0}.
\end{align}
Similarly,
\begin{equation}
  \Gamma_4 = -i\bra{M_0} e^{i\ham_0 t} c_3 e^{-i \ham ' t} c_6 \ket{M_0},
\end{equation}
from which we finally obtain Eq.~(\ref{eq:gammas}) after a time integration using the Lehmann spectral representation.
\section{Bogoliubov transformations}

We introduce complex matter fermions on each $z$ bond $\langle ij\rangle_z$ ($i \in A, j \in B$) at position $r$, and we relabel the Majorana fermions as $c_i = c_{Ar}$, $c_j = c_{Br}$, such that
\begin{equation}
  f_r = \frac{c_{Ar} + ic_{Br}}{2}, \qquad  f_r\dag = \frac{c_{Ar} - ic_{Br}}{2}.
\end{equation}

The Hamiltonians $\ham_0$ and $\ham'$ can then be diagonalized on a finite system. The resulting complex fermions $a\dag$ and $b\dag$ of Eq.~(\ref{eq:diagham}) and the $f\dag$ fermions are related by a Bogoliubov transformation
\begin{equation}
\label{eq:bogo}
\begin{pmatrix}
	a \\ a\dag 
\end{pmatrix}
=
\begin{pmatrix}
    X_a^* & Y_a^* \\
    Y_a & X_a
\end{pmatrix}
\begin{pmatrix}
	f \\ f\dag 
\end{pmatrix}
,\qquad
\begin{pmatrix}
	b \\ b\dag 
\end{pmatrix}
=
\begin{pmatrix}
    X_b^* & Y_b^* \\
    Y_b & X_b
\end{pmatrix}
\begin{pmatrix}
	f \\ f\dag 
\end{pmatrix}
,\qquad
\begin{pmatrix}
	b \\ b\dag 
\end{pmatrix}
=
\begin{pmatrix}
    \mathcal{X}^* & \mathcal{Y}^* \\
    \mathcal{Y} & \mathcal{X}
\end{pmatrix}
\begin{pmatrix}
	a \\ a\dag 
\end{pmatrix},
\end{equation}
where $\mathcal{X}$ and $\mathcal{Y}$ are related to $X_{a,b}$ and $Y_{a,b}$ \cite{blaizot1986quantum, knolle2016dynamics}. We use the notation $f\dag = (f\dag_1 \dots f\dag_N)$, and similarly for column vectors.
As the ground states of $\ham_0$ and $\ham'$, $\ket{M_0}$ and $\ket{M'}$ respectively, have the same parity, $\braket{M_0}{M'} \neq 0$ and
\begin{equation}
\label{eq:GSrel}
  \ket{M'} = [\det(\mathcal{X} \dag \mathcal{X})]^{1/4} e^{-\frac{1}{2} \sum_{nm}\mathcal{F}_{nm} a\dag_n a\dag_m} \ket{M_0},
\end{equation}
where $\mathcal{F} = \mathcal{X}^{*-1}\mathcal{Y}$ \cite{blaizot1986quantum}. For Hamiltonians with different ground state parities, $\mathcal{X}$ is singular and such expression does not exist.
For single-particle eigenstates $\ket{\lambda} = b\dag_{\lambda} \ket{M'}$ of $\ham'$ we find
\begin{align}
	\mel{M_0}{c_{Ar} b\dag_{\lambda} }{M'} &= \, \,[\det(\mathcal{X} \dag \mathcal{X})]^{1/4} \left[ (\mathcal{X}\dag)^{-1} ( X_a + Y_a )\right]_{\lambda r} \nonumber \\
	\mel{M_0}{c_{Br} b\dag_{\lambda} }{M'} &= i[\det(\mathcal{X} \dag \mathcal{X})]^{1/4} \left[ (\mathcal{X}\dag)^{-1} ( Y_a - X_a )\right]_{\lambda r},
\end{align}
and for three-particle eigenstates $\ket{\lambda} = b\dag_{\lambda_1} b\dag_{\lambda_2} b\dag_{\lambda_3} \ket{M'}$, we find
\begin{alignat}{3}
	\mel{M_0}{c_{Ar} b\dag_{\lambda_1} b\dag_{\lambda_2} b\dag_{\lambda_3} }{M'} & &= \, \,[\det(\mathcal{X} \dag \mathcal{X})]^{1/4} \Big\{  & \qty[ (\mathcal{X}\dag)^{-1} ( X_a + Y_a )]_{\lambda_1 r} \qty[ (\mathcal{Y}(\mathcal{X}^*)^{-1}]_{\lambda_2 \lambda_3}\nonumber \\
	&  &+ & \qty[ (\mathcal{X}\dag)^{-1} ( X_a + Y_a )]_{\lambda_2 r} \qty[ (\mathcal{Y}(\mathcal{X}^*)^{-1}]_{\lambda_3 \lambda_1}\nonumber \\
	& & + & \qty[ (\mathcal{X}\dag)^{-1} ( X_a + Y_a )]_{\lambda_3 r} \qty[ (\mathcal{Y}(\mathcal{X}^*)^{-1}]_{\lambda_1 \lambda_2} \Big\}\nonumber \\
	\mel{M_0}{c_{Br} b\dag_{\lambda_1} b\dag_{\lambda_2} b\dag_{\lambda_3} }{M'} & & = \, \,i[\det(\mathcal{X} \dag \mathcal{X})]^{1/4} \Big\{  & \qty[ (\mathcal{X}\dag)^{-1} ( Y_a - X_a )]_{\lambda_1 r} \qty[ (\mathcal{Y}(\mathcal{X}^*)^{-1}]_{\lambda_2 \lambda_3}\nonumber \\
	&  &+ & \qty[ (\mathcal{X}\dag)^{-1} ( Y_a - X_a )]_{\lambda_2 r} \qty[ (\mathcal{Y}(\mathcal{X}^*)^{-1}]_{\lambda_3 \lambda_1}\nonumber \\
	& & + & \qty[ (\mathcal{X}\dag)^{-1} ( Y_a - X_a )]_{\lambda_3 r} \qty[ (\mathcal{Y}(\mathcal{X}^*)^{-1}]_{\lambda_1 \lambda_2} \Big\}\nonumber \\.
\end{alignat} 

\section{Symmetries of $\ham_0$ and $\ham'$}
Let $\bm{r}_0$ be the center of the $\langle 25 \rangle$ bond depicted in Fig.~\ref{fig:fixhexagons} of the main text. Then, to each site $i$ corresponds a site $p(i)$ such that $(\bm{r}_i - \bm{r}_0) = - (\bm{r}_{p(i)} - \bm{r}_0)$ modulo the periodic boundary conditions. If $i \in A$ then $p(i) \in B$ and vice versa. The transformation $p$ is bijective and $p^{-1} = p$.
The Hamiltonian $\ham_0$ and $\ham'$ are such that for all bonds $\langle ij \rangle_{\gamma}$,
\begin{equation}
\label{eq:parity}
  u_{\langle ij \rangle_{\gamma}} = u_{\langle p(j) p(i) \rangle_{\gamma}}.
\end{equation}

We then define the unitary transformation $U_I$ such that 
\begin{equation}
  U_I c_i U_I\dag = c_{p(i)},
\end{equation}
for all $i \in A, B$, and the particle-hole unitary transformation $U_{PH}$,
\begin{alignat}{2}
  U_{PH} c_i U_{PH}\dag &= c_i \quad & & \forall i \in A \nonumber \\
  U_{PH} c_j U_{PH}\dag &= -c_j \quad & & \forall j \in B.
\end{alignat}
It corresponds to a particle-hole symmetry in the sense that $U_{PH} f\dag U_{PH}\dag = f$ and $U_{PH} f U_{PH}\dag = f\dag$.
Finally, we define the combined unitary transformation $V = U_{PH} U_I$, which satisfies
\begin{alignat}{2}
  V c_i V\dag &= -c_{p(i)} &\quad &\forall  i \in A \nonumber \\
  V c_j V\dag &= c_{p(j)}  &\quad &\forall j \in B,
\end{alignat}
so that $V^2 c_i (V\dag)^2= -c_i, \, \, \forall i$.
Additionally, for any matter Hamiltonian $\ham$ satisfying Eq.~(\ref{eq:parity}),
\begin{equation}
  \{\ham, U_I\} = 0, \quad \{\ham, U_{PH}\} = 0, \quad [\ham, V] =0.
\end{equation}

We can therefore choose a basis of eigenstates of $\ham$ so that each element $\ket{\lambda}$ satisfies
\begin{align}
  \ham \ket{\lambda} = E_{\lambda} \ket{\lambda}, \qquad
  V \ket{\lambda} = v_{\lambda} \ket{\lambda}.
\end{align}
Using the properties of $V$, the following eigenstates of $V$ can be constructed from any $\ket{\lambda}$,
\begin{align}
\label{eq:dfermions}
  \ket{\lambda_+} &= d_m\dag \ket{\lambda}, \nonumber \\
  \ket{\lambda_-} &= d_m \ket{\lambda}, \nonumber \\
  V \ket{\lambda_{\pm}} &= \mp i v_{\lambda} \ket{\lambda_{\pm}},
\end{align}
where we have defined the complex fermions
\begin{align}
  d_m\dag = \frac{c_m + i c_{p(m)}}{2}, \quad d_m = \frac{c_m - i c_{p(m)}}{2}, \quad {\rm for} \, \, m\in A.
\end{align}

Then using a Bogoliubov transformation relating the $d\dag$ fermions to the $b\dag$ fermions, and a series of arguments, we argue that we can sort the fermions into two species: $b\dag = ( b_+\dag \,\, b_-\dag )$, which satisfy
\begin{equation}
  V b\dag_+ V\dag = -i b\dag_+ , \qquad  V b\dag_- V\dag = ib\dag_-.
\end{equation}
Thanks to Eq.~(\ref{eq:GSrel}), we can also show that $\ket{M_0}$ and $\ket{M'}$ have the same $V$ eigenvalue:  $V \ket{M_0} = v_0 \ket{M_0}$ and $V \ket{M'} = v_0 \ket{M'}$. Note that we have assumed that $\omega_n \neq 0$ and $\omega'_n \neq 0$ for all $n$.

Finally, for $j\in B$,
\begin{align}
  \mel{M_0}{c_j}{\lambda} &= \mel{M_0}{V\dag  c_{p(j)} V}{\lambda} \nonumber \\
  &= \frac{1}{v_0} \mel{M_0}{c_{p(j)} V b\dag_{\lambda_1} \dots b\dag_{\lambda_m}}{M'}, \quad (m{\rm \, \, \, odd}) \nonumber \\
  &= \pm i \mel{M_0}{c_{p(j)}}{M'},
\end{align}
where the sign depends on the composition of $\ket{\lambda}$ in terms of $b\dag_+$ and $b\dag_-$ fermions.
Moreover, for single-particle states $\ket{\lambda} = b\dag_{\lambda} \ket{M'}$ we have
\begin{equation}
  \mel{M_0}{c_j b\dag_{\lambda} }{M'} = \left\{
  	\begin{array}{ll}
  	-i \mel{M_0}{c_{p(j)} b\dag_{\lambda} }{M'} , \quad {\rm if \, \,} b_{\lambda} \in \{b_+ \}  \\
  	\, \, \, \, \, i \mel{M_0}{c_{p(j)} b\dag_{\lambda} }{M'} , \quad {\rm if \, \,} b_{\lambda} \in \{ b_- \} 
    \end{array}
  \right.
\end{equation}
For $j\in A$, the signs are reversed.

In the optical conductivity, only expressions of the form $\mel{M_0}{c_j b\dag_{\lambda} }{M'} - i\mel{M_0}{c_{p(j)} b\dag_{\lambda} }{M'}$ with $j\in B$ appears so that only $b\dag_-$ states contributes. In the magnetic susceptibility however, only $b\dag_+$ states contribute. This can be generalized to any odd-particle energy eigenstates.

\end{document}